\documentclass[conference]{IEEEtran}

\pdfoutput=1

\newlength{\figwidth}
\usepackage{ifthen} 
\usepackage{substr}

\usepackage{color}
\definecolor{links}{rgb}{0.7,0,0}   
\definecolor{urls}{rgb}{0,0,0.8}    
\definecolor{cites}{rgb}{0,0,0.8}   

\usepackage{tikz}
\usetikzlibrary{arrows}

\usepackage[nosort]{cite}
\usepackage{url}
\usepackage[intlimits]{amsmath}
\usepackage{bbm}
\usepackage{graphicx}
\usepackage{paralist}
\usepackage{fancyref}
\usepackage[stretch=16,shrink=16,step=4]{microtype}
\usepackage[inline]{enumitem}

\usepackage{header}

\usepackage{kolor}

\usepackage{kontext}

\usepackage{konfig}

\let\marginnote\undefined

\newcommand{\marginnote}[1]{\marginpar{\framebox{\framebox{#1}}}}

\begin{document}

\IEEEoverridecommandlockouts

\title{Delay and Peak-Age Violation Probability \\  in Short-Packet Transmissions}
\author{\IEEEauthorblockN{Rahul Devassy$^1$, Giuseppe Durisi$^1$,  Guido Carlo Ferrante$^1$,  Osvaldo Simeone$^2$, and  Elif Uysal-Biyikoglu$^3$}\\[-3mm]
\IEEEauthorblockA{
$^1$Chalmers University of Technology, 41296 Gothenburg, Sweden\\$^2$King's College London, London, WC2R 2LS, United Kingdom\\$^3$Middle East Technical University, Ankara, Turkey}
\thanks{This work was partly funded by the Swedish Research Council under grant 2012-4571.
The simulations were performed in part on resources provided by the Swedish National Infrastructure for Computing (SNIC)  at C3SE.}
\thanks{Osvaldo Simeone has received funding from the European Research Council (ERC) under the European Union’s Horizon 2020 Research and Innovation Programme (Grant Agreement No. 725731).}
\thanks{Elif Uysal-Biyikoglu has received funding from T\"{U}B\.{I}TAK (Grant No. 117E215).}
}
%
\maketitle

\begin{abstract}
This paper investigates the distribution of delay and peak age of information in a communication system where packets, generated according to an independent and identically distributed Bernoulli process, are placed in a single-server queue with first-come first-served discipline and transmitted over an additive white Gaussian noise (AWGN) channel.
When a packet is correctly decoded, the sender receives an instantaneous error-free positive acknowledgment, upon which it removes the packet from the buffer.
In the case of negative acknowledgment, the packet is retransmitted.
By leveraging finite-blocklength results for the AWGN channel, we characterize the delay violation and the peak-age violation probability without resorting to approximations based on large deviation theory as in previous literature.
Our analysis reveals that there exists an optimum blocklength that minimizes the delay violation and the peak-age violation probabilities.
We also show that one can find two blocklength values that result in very similar average delay but significantly different delay violation probabilities.
This highlights the importance of focusing on violation probabilities rather than on averages.
\end{abstract}
\section{Introduction}
\label{sec:introduction}
Emerging wireless applications such as factory automation and vehicular communication require the availability of mission-critical links that are able to deliver short information packets within stringent latency and reliability constraints.
As shown in~\cite{polyanskiy2010channel}, finite-blocklength information theory provides accurate tools to describe the tradeoff between latency, reliability, and rate when transmitting short packets. %
Leveraging tools from non-asymptotic information theory, the purpose of this paper is to analyze the probability that the delay or the peak age~\cite[Def.~3]{costa2016age} exceeds a predetermined threshold in a point-to-point communication system with random information-packet arrivals per channel use.
The analysis assumes a single-server queue operating according to a first-come-first-served (FCFS) policy.

\paragraph*{Related Work} 
Aside from the work by Telatar and Gallager~\cite{telatar1995combining}, who employed an error-exponent approach, most  works in queuing analysis of communication links rely on a bit-pipe abstraction of the physical layer.
Accordingly, bits are delivered reliably at a rate equal to the channel capacity, or in the case of fading channels, at a rate equal to the outage capacity for a given outage probability.
These works may be classified into three broad categories:
\begin{enumerate*}[label=(\roman*)]
\item analyses of the steady-state average delay;
\item analyses of the delay violation probability using large deviation theory (see~\cite{yeh12-a,al2016network} and references therein); and
\item analyses of throughput-delay tradeoff under deadline constraints~\cite{zafer2009minimum,singh2015decentralized}. \end{enumerate*}
However, the bit-pipe abstraction is not suitable when the latency constraints prevent the use of channel codes with long blocklength.
Indeed, outage and ergodic capacity are poor performance benchmarks when packets are short~\cite{durisi16-09a}, and using them may result in inaccurate delay estimates.

Recognizing these limitations, Hamidi-Sepehr \emph{et al.}~\cite{hamidi2015delay} analyzed the queuing behavior when BCH codes are used.
Specifically, they evaluated both the probability distribution of the steady-state queue size and the average delay.
A different approach, which relies on random coding, is to replace capacity or outage capacity with the more accurate second-order asymptotic approximations obtained in~\cite{polyanskiy2010channel,yang14-07c}.
This approach has been used in~\cite{Gursoy2013} to study the throughput achievable over a fading channel under a constraint on the probability of buffer overflow; in~\cite{schiessl2015delay} to analyze the packet delay violation probability in the presence of perfect channel-state information at the transmitter, which allows for rate adaptation; and in~\cite{she16-12b} to design the downlink of an ultra-reliable transmission system under a constraint on the end-to-end delay.
All these works rely on large-deviation theory through effective capacity~\cite{wu2003effective}, stochastic network calculus~\cite{jiang2008stochastic}, and effective bandwidth~\cite{chang1995effective}, hence providing tight delay estimates only in the asymptotic limit of large delay.

For applications in which packets carry status updates, the time elapsed since the newest update available at the destination was generated at the source, commonly referred to as age of information, is more relevant than delay.
Most previous analyses of the age of information focus on its average or peak value (see~\cite{costa2016age} and references therein), and rely on simple physical-layer models.
A recent exception is~\cite{inoue2017stationary}, where the stationary distribution of the peak age is characterized, and~\cite{sun2017update}, where generalized age penalty functions are analyzed.

\paragraph*{Contributions} 
We analyze the delay and peak-age violation probabilities achievable over an additive white Gaussian noise (AWGN) channel where information packets arrive in each channel use (CU) according to an independent and identically distributed (\iid) Bernoulli process and are transmitted using an FCFS policy with automatic repeat request.
Our specific contributions are as follows:
\begin{itemize}[leftmargin=*]
\item We determine in closed form the probability-generating functions (PGFs) of  delay and peak age at steady state.
Delay and peak-age violation probabilities can be efficiently obtained from the derived PGFs through an inverse transform.
We also present an accurate approximation of this inverse transform based on saddlepoint methods.
\item We numerically illustrate the dependence of delay on the  blocklength.
Specifically, we show that there exist two  blocklength values resulting in the same average delay, but yielding delay violation probabilities that differ by two orders of magnitude.
This shows that average delay is insufficient in capturing performance.%
\item Finally, we discuss the accuracy of delay violation estimates based on the large-deviation tools used in~\cite{schiessl2015delay}. 
\end{itemize}

\paragraph*{Notation}
Uppercase boldface letters denote random quantities and lightface letters denote deterministic quantities.
The distribution of a random variable~$\chinp$ is denoted by~$P_\chinp.$
With~$\Bexpectation{\cdot}$ we denote the expectation operator.
The indicator function and the ceil function are denoted by~$\indicator{\cdot}$ and~$\ceil{\cdot}$ respectively.
We let~$\Bbernoullidist{p}$ denote a Bernoulli-distributed random variable with parameter~$p,$~$\Bbinomialdist{n}{p}$ denote a Binomial-distributed random variable with parameters~$n$ and~$p,$  and~$\Bgeometricdist{p}$ a geometrically distributed random variable with parameter~$p.$
The PGF of a nonnegative integer-valued random variable~$\chinp$ is $G_{\chinp}(s)= \Bexpectation{s^\chinp}$.

\section{Frame-Synchronous Model}\label{sec:frame_sync_sys_mod}
We consider a point-to-point discrete-time AWGN channel.
The information-packet arrival process is \iid Bernoulli over the CUs.
Specifically, the probability of a new packet arrival in each CU is $\packetarrivalprob.$
The information packets are stored at the transmitter in a single-server queue operating according to an FCFS policy.
Each information packet consists of~$\infobits$ bits, which are mapped into a codeword of blocklength $n$ CUs and power~$\snr$.\footnote{We assume that the variance of the Gaussian additive noise is one. So~$\snr$ is also the signal-to-noise ratio.}
The packet error probability is denoted by $\errorprob>0$.
A packet is removed from the buffer when its reception is acknowledged by the receiver through an  ACK feedback.
If the codeword is not correctly decoded, the receiver sends a NACK message and the codeword is retransmitted.
We assume perfect error detection at the receiver and instantaneous error-free ACK/NACK transmission, as commonly done in the literature.

We will first assume that time is organized into time frames of duration $\blocklength$ CUs so that the transmission of a codeword can only start at the beginning of a time frame.
Under this assumption, if an information packet arrives when the buffer is empty, its transmission is scheduled for the next available frame.
We refer to this setup as being \emph{frame synchronous}.
In Section~\ref{sec:relaxing_the_frame_synchronism}, we shall relax this assumption and allow transmission to start in the next available CU when the buffer is empty.
We refer to this setup as being \emph{frame asynchronous}.
This model yields a reduction in latency at the cost of a more involved frame-synchronization procedure.
Under the frame-synchronous assumption, the system can be modeled as a $\mathit{Geo}/G/1$ queue with bulk arrivals, sometimes denoted $\mathit{Geo}^{[X]}/G/1$ (see \cite[Sec. 4.6.2]{bose2013introduction}).

We group together all packets arriving within a time frame as a \emph{bulk}, and study the evolution of the transmitter's buffer along the time index $t$ running over the time frames.
Let $\bulkarrivalat{t}$ be the number of packets received in the $t$-th time frame.
It follows that the process $\curlybrac{\bulkarrivalat{t}}_{t=1}^{\infty}$ is \iid with $\Bbinomialdist{\blocklength}{\packetarrivalprob}$ marginal distribution.
When $\bulkarrivalat{t}>0$, we say that a bulk has been received at time frame $t$.
Furthermore, we denote by $\queuesizeattime{t}$ the number of bulks remaining in queue at the start of the $(t+1)$th time frame.

Let $\arrivaltimeofbulk{m}$ be the frame index corresponding to the arrival time of the  $m$th bulk, and $\bulkarrivalcountat{m}$ be the number of packets in the $m$th bulk.
Note that $\{\bulkarrivalcountat{m}\}_{m=1}^{\infty}$ is an \iid process with marginal distribution equal to the conditional distribution of $\bulkarrivalat{t}$ given the event $\curlybrac{\bulkarrivalat{t}>0}.$ %
We denote by~$\waitingtimeofbulk{m}$ the waiting time of the~$m$th bulk, i.e.,  the number of frames the first packet in the bulk remains in the queue before being served.
Moreover, $\servicetimeofbulk{m}$ is the service time of the~$m$th bulk, i.e., the total number of frames needed to successfully transmit all packets in the~$m$th bulk.
The service process $\curlybrac{\servicetimeofbulk{m}}_{m=1}^{\infty}$ is \iid with 
\begin{IEEEeqnarray}{rCl}
\servicetimeofbulk{m} &\sim& \sum_{k=1}^{\bulkarrivalcountat{m}}\servicetimeofpacket{k}\label{def:servicetime_bulk}
\end{IEEEeqnarray}
where the variables $\curlybrac{\servicetimeofpacket{k}}_{k=1}^{\bulkarrivalcountat{m}}$ are \iid  $\Bgeometricdist{1-\errorprob}$-distributed and independent of~$\bulkarrivalcountat{m}.$
Each variable~$\servicetimeofpacket{k}$ represents the number of time frames needed to reliably deliver one packet.
Finally, the delay $\delayat{m}=\waitingtimeofbulk{m}+\servicetimeofbulk{m}$ of the $m$th bulk (measured in frames) is the sum of waiting time $\waitingtimeofbulk{m}$ and service time $\servicetimeofbulk{m}.$
For this queuing system, the process~$\curlybrac{\delayat{m}}_{m=1}^{\infty}$ has a steady-state distribution as long as~$\packetarrivalprob\blocklength<1-\errorprob$.
This distribution is studied in the next section.
We will discuss the peak-age metric in Section~\ref{sec:peak_age_of_information}.

\section{Steady-State Delay Violation Probability}\label{sec:steady-state}

In this section, we focus on the analysis of the steady-state delay violation probability.
This is defined as
\begin{align}
P_{\mathrm{dv}}(\delaythreshold)  = \lim_{m\to\infty} \probof{\delay_{m}\geq\delaythreshold/\blocklength}
 				 =\probof{\delay\geq\delaythreshold/\blocklength}, \label{expr:delayviolation_criteria}
\end{align}
where~$\delay$ is the steady-state delay and $\delaythreshold$ is the desired latency constraint (measured in CUs).
To characterize $P_{\mathrm{dv}}(d_{0})$, we will first derive the PGF of~$\delay$, and then obtain~$P_{\mathrm{dv}}(d_{0})$ implicitly through an inversion formula.
As the PGF of the delay~$\delay$ for our setup is not directly available in the literature, although its derivation follows along the steps described in~\cite[Sec. 4.6.2]{bose2013introduction}, we provide it in the following theorem.
\begin{thm}\label{thm:delay_steadystate}
For every $\epsilon>0$ such that $\packetarrivalprob\blocklength<1-\errorprob$, the PGF of the steady-state delay $\delay$ for the frame-synchronous model is 
\vspace{-0.0pt}
\begin{multline}
 G_{\delay}(s) = \parantheses{1-\packetarrivalprob\blocklength/(1-\errorprob)}   \\  \cdot\frac{\parantheses{1-s}\parantheses{\parantheses{1-\packetarrivalprob}^\blocklength\parantheses{1-\errorprob s}^\blocklength-\parantheses{1-\packetarrivalprob+\parantheses{\packetarrivalprob-\errorprob}s}^\blocklength}} {\parantheses{1-\parantheses{1-\packetarrivalprob}^\blocklength}\parantheses{s\parantheses{1-\errorprob s}^\blocklength-\parantheses{1-\packetarrivalprob+\parantheses{\packetarrivalprob-\errorprob}s}^\blocklength}}.\label{expr:pgf_steadystate_delay}
\end{multline}
\end{thm}
\begin{IEEEproof}
See Appendix~\ref{proof:delay_steadystate}.
\end{IEEEproof}

The delay violation probability~\eqref{expr:delayviolation_criteria} can be obtained from~\eqref{expr:pgf_steadystate_delay} through the following inversion formula
\begin{IEEEeqnarray}{rCl}
P_{\mathrm{dv}}(\delaythreshold) &=&  1- \parantheses{\frac{1}{2\pi i}\oint\nolimits_{C} \frac{G_{\delay}\parantheses{s}}{(1-s)s^{d-1}}\infinitesimal s}\indicator{d\geq 2}
\IEEEeqnarraynumspace \label{expr:ccdf_steadystate_delay}
\end{IEEEeqnarray}
where~$d=\left\lceil\delaythreshold/\blocklength\right\rceil$ and~$C$ is a circle centered at the origin of the complex plane enclosing all poles of $G_{\delay}(s)/(1-s).$
Since the contour integral in~\eqref{expr:ccdf_steadystate_delay} is not known in closed form,
the numerical evaluations of $P_{\mathrm{dv}}(\delaythreshold)$ we shall present in Section~\ref{sec:numerical_results} are based on a recursion-based $z$-transform inversion~\cite[Eq. (10)]{jenkins1967useful} of $G_{\delay}(s)/(1-s)$.

A reduced-complexity approach to compute the delay violation probability from~\eqref{expr:pgf_steadystate_delay} is through the saddlepoint method~\cite[Eq. (2.2.10)]{jensen1995saddlepoint}, which, under the assumption
 that~$\left\lceil\delaythreshold/\blocklength\right\rceil> \Bexpectation{\delay}=\lim_{s\uparrow 1}G_{\delay}'\parantheses{s}$, results in the following approximation\footnote{See~\cite[p.~27]{jensen1995saddlepoint} for an extension to the case $\left\lceil\delaythreshold/\blocklength\right\rceil< \Bexpectation{\delay}$.}
\begin{IEEEeqnarray}{rCl}
P_{\mathrm{dv}}(\delaythreshold) &\approx&  \frac{B_0\parantheses{\theta\sigma\parantheses{\theta}}}{\sigma\parantheses{\theta}\parantheses{1-e^{-\theta}}} e^{\kappa\parantheses{\theta}-\theta \left\lceil\delaythreshold/\blocklength\right\rceil}. \label{expr:ccdf_steadystate_delay_saddlepoint}
\end{IEEEeqnarray}
In~\eqref{expr:ccdf_steadystate_delay_saddlepoint}, %
$\kappa\parantheses{x} = \Blog{G_{\delay}\parantheses{e^x}}$, %
$\theta = \argmin_{x\in\mathbb{R}} \kappa\parantheses{x}-x\left\lceil\delaythreshold/\blocklength\right\rceil$, %
$\sigma\parantheses{x} =  \sqrt{\kappa''\parantheses{x}}$, and %
$B_0\parantheses{x} = xe^{x^2/2}Q\parantheses{x}$, %
where~$Q\parantheses{x}$ is the Gaussian Q-function and the prime notation denotes derivatives.

We present next, for comparison purposes, an upper bound on~\eqref{expr:delayviolation_criteria} obtained through a stochastic-network calculus approach~\cite{jiang2008stochastic}.
The proof of this bound, which is easier to evaluate than~\eqref{expr:ccdf_steadystate_delay} but less tight, involves specializing the general result reported in~\cite[Thm. 1]{al2016network} to our setup.
\begin{thm}\label{thm:ccdf_steadystate_delay_netcalc}
The delay violation probability $P_{\mathrm{dv}}(\delaythreshold) $ in~\eqref{expr:delayviolation_criteria} is upper-bounded as
\begin{equation}
P_{\mathrm{dv}}(\delaythreshold)  \leq \hspace{-2ex} \inf_{\substack{s>1:\\G_{\bulkarrivalat{}}({s})G_{\servicedbitsperslot}({1/s})<1}} \frac{G_{\servicedbitsperslot}\parantheses{1/s}^{d-1}}{1-G_{\bulkarrivalat{}}\parantheses{s}G_{\servicedbitsperslot}\parantheses{1/s}},\label{expr:ccdf_steadystate_delay_netcalc}
\end{equation}
where~$d=\left\lceil\delaythreshold/\blocklength\right\rceil$ and the PGFs $G_{\bulkarrivalat{}}\parantheses{s}$ and $G_{\servicedbitsperslot}\parantheses{s}$ are
\begin{equation}
G_{\bulkarrivalat{}}\parantheses{s} = \parantheses{1-\packetarrivalprob+\packetarrivalprob s}^\blocklength, \quad
G_{\servicedbitsperslot}\parantheses{s} = \errorprob + \parantheses{1-\errorprob} s.
\end{equation}
\end{thm}
\begin{IEEEproof}
By following the analysis in~\cite[Sec. 4.4]{schiessl2015delay}, we have
\begin{IEEEeqnarray}{rCl}
P_{\mathrm{dv}}(\delaythreshold)
&\leq& \lim_{t\tendsto\infty}\inf_{s>1} \sum\nolimits_{u=0}^{t} G_{\bulkarrivalat{}}\parantheses{s}^{t-u} G_{\servicedbitsperslot}\parantheses{1/s}^{t+d-1-u}\label{step:usestochasticnetcalresultfordelayviolation1}\\
&\leq& \inf_{s>1} G_{\servicedbitsperslot}\parantheses{1/s}^{d-1} \sum\nolimits_{u=0}^{\infty} G_{\bulkarrivalat{}}\parantheses{s}^{u} G_{\servicedbitsperslot}\parantheses{1/s}^{u}.\IEEEeqnarraynumspace\label{step:usestochasticnetcalresultfordelayviolation2}
\end{IEEEeqnarray}
Here,~\eqref{step:usestochasticnetcalresultfordelayviolation1} follows from~\cite[Eqs.~(21)--22)]{schiessl2015delay}.
We obtain~\eqref{expr:ccdf_steadystate_delay_netcalc} by computing the geometric series in~\eqref{step:usestochasticnetcalresultfordelayviolation2}.
\end{IEEEproof}

\section{Asynchronous Model and Analysis} 
\label{sec:relaxing_the_frame_synchronism}
We consider a variation of the setup described in Section~\ref{sec:frame_sync_sys_mod}, in which, if the buffer is empty when a packet arrives, the corresponding codeword is transmitted starting from the next available CU.
We  refer to this model as being frame asynchronous. The rationale for this terminology is that, in this setup, frame synchronization between transmitter and receiver needs to be reacquired whenever the buffer is empty.

Since in the frame-asynchronous case packets are not grouped into bulks, this  setup can be modeled as a~$Geo/G/1$ queue.
The PGF of the steady-state delay $\delayasync$ measured in CUs is given in the following theorem, whose proof follows along the same lines as the proof of Theorem~\ref{thm:delay_steadystate}.
\begin{thm}\label{thm:ccdf_steadystate_delay_async}
For every $\epsilon>0$ such that $\packetarrivalprob\blocklength<1-\errorprob$, the PGF of the steady-state delay for the frame-asynchronous model is
\begin{IEEEeqnarray}{rCl}
 G_{\delayasync}(s) &=& \frac{\parantheses{s-1}(1-\errorprob-\packetarrivalprob\blocklength)s^\blocklength}{s-(1-\packetarrivalprob)-(\packetarrivalprob+\errorprob(s-1))s^\blocklength}.\label{expr:pgf_steadystate_delay_async}
\end{IEEEeqnarray}
\end{thm}

The delay violation probability and its saddlepoint approximation can be obtained  by proceeding as in Section~\ref{sec:steady-state}.
However, differently from Section~\ref{sec:steady-state}, we cannot obtain a stochastic-network calculus upper bound similar to the one in Theorem~\ref{thm:ccdf_steadystate_delay_netcalc}.
Indeed, in the frame-asynchronous setup, the independence assumption made in~\cite[Lem. 4]{al2016network}, which is needed in proof of the delay violation probability upper bound~\cite[Thm. 1]{al2016network}, is violated.%

\section{Steady-State Peak-Age Violation Probability} 
\label{sec:peak_age_of_information}

\begin{figure}[t]
  \centering
\begin{tikzpicture}[font=\footnotesize,scale=0.55]
	\draw[-latex'] (0,0) -- (0,7);
	\draw[-latex'] (0,0) -- (10,0);
	\draw (10,0) node[below]{Time};
	\draw (0,7) node[left]{Age};
	\draw[thick] (0,1) -- (3,4) -- (3, 1) -- (8,6) -- (8,3) -- (10,5);
	\draw[gray,dashed] (3,1) -- (2,0);
	\draw[gray,dashed] (3,1) -- (3,0);
	\draw[gray,dashed] (8,3) -- (5,0);
	\draw[gray,dashed] (8,3) -- (8,0);
	\draw (2,0) node[below]{$\arrivaltimeofbulk{1}\hspace{1ex}$};
	\draw (3,0) node[below]{$\hspace{2ex} \arrivaltimeofbulk{1}\!+\!\delay_{1}$};
	\draw (5,0) node[below]{$\arrivaltimeofbulk{2}$};
	\draw (8,0) node[below]{$\arrivaltimeofbulk{2}+\delay_{2}$};
	\draw (3,4) node[above]{$\peakageofinfocountat{1}$};
	\draw (8,6) node[above]{$\peakageofinfocountat{2}$};
\end{tikzpicture}
\caption{Peak age of information for the frame-synchronous model: $\arrivaltimeofbulk{m}$ is the frame index corresponding to the arrival of bulk $m$; $\arrivaltimeofbulk{m}+\delay_{m}$ is the frame index corresponding to its departure; the peak age $\peakageofinfocountat{m}$ is the age of information just before the $m$th bulk departs.}\label{fig:peak_age}
\end{figure}
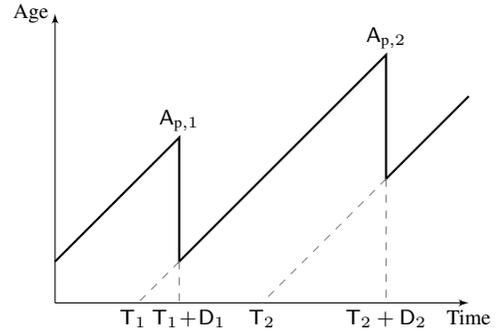

We next characterize the violation probability of the steady-state peak age for both the frame-synchronous and frame-asynchronous models.

The peak-age of information is the  value of the age of information just before an update is received (see Fig.~\ref{fig:peak_age}).
The peak age $\peakageofinfocountat{m}$ can be written as~\cite[Eq. (9)]{inoue2017stationary}
\begin{equation}\label{eq:peak_age} \peakageofinfocountat{m}=\max\{\delay_{m-1},\arrivaltimeofbulk{m}-\arrivaltimeofbulk{m-1}\}+\servicetimeofbulk{m}.
\end{equation}
In words, it is the sum of the service time of the~$m$th bulk~$\servicetimeofbulk{m}$  and the maximum between the delay~$\delay_{m-1}$ of the $(m-1)$th bulk of packets  and the difference~$\arrivaltimeofbulk{m}-\arrivaltimeofbulk{m-1}$ between the frame indices corresponding to the arrival of the $m$th and the $(m-1)$th bulks.
%
The PGF of the steady-state peak age $\peakageofinfo= \max\parantheses{\delay,\arrivaltimeofbulk{2}-\arrivaltimeofbulk{1}} + \servicetimeofbulk{1}$ can be derived similarly as in~\cite[Thm. 9]{inoue2017stationary} and is given in the next theorem.
\begin{thm}\label{thm:pgf_steadystate_peakage}
For every $\epsilon>0$ such that $\packetarrivalprob\blocklength<1-\errorprob$, the PGF of the peak age of information $\peakageofinfo$ at steady state for the frame-synchronous model is
\begin{IEEEeqnarray}{rCl}
G_{\peakageofinfo}(s) &=& \frac{ \parantheses{1-\packetarrivalprob +\parantheses{\packetarrivalprob-\errorprob}s}^\blocklength - \parantheses{1-\errorprob s}^\blocklength \parantheses{1-\packetarrivalprob}^\blocklength }{\parantheses{1-\errorprob s}^\blocklength\parantheses{1-\parantheses{1-\packetarrivalprob}^\blocklength}} \nonumber\\
&&. \parantheses{G_{\delay}(s)-\frac{\parantheses{1-s}G_{\delay}(\parantheses{1-\packetarrivalprob}^\blocklength s)}{1-\parantheses{1-\packetarrivalprob}^\blocklength s}} \IEEEeqnarraynumspace\label{expr:pgf_steadystate_peakage}
\end{IEEEeqnarray}
where~$G_{\delay}(s)$ is given in~\eqref{expr:pgf_steadystate_delay}.
For the frame-asynchronous model the steady state peak age of information is given as
\begin{IEEEeqnarray}{rCl}
G_{\peakageofinfo}(s) &=& \frac{\parantheses{1-\errorprob}s^\blocklength}{1-\errorprob s^\blocklength} \parantheses{G_{\delay}(s)-\frac{\parantheses{1-s}G_{\delay}(\parantheses{1-\packetarrivalprob} s)}{1-\parantheses{1-\packetarrivalprob} s}} \IEEEeqnarraynumspace\label{expr:pgf_steadystate_peakage_async}
\end{IEEEeqnarray}
where~$G_{\delay}(s)$ is given in~\eqref{expr:pgf_steadystate_delay_async}.
\end{thm}

The peak-age violation probability and the corresponding saddlepoint approximation for both cases can be obtained by proceeding as in Section~\ref{sec:steady-state}.
%
%
\begin{figure}[t]
\centering
\includegraphics[width=0.9\columnwidth]{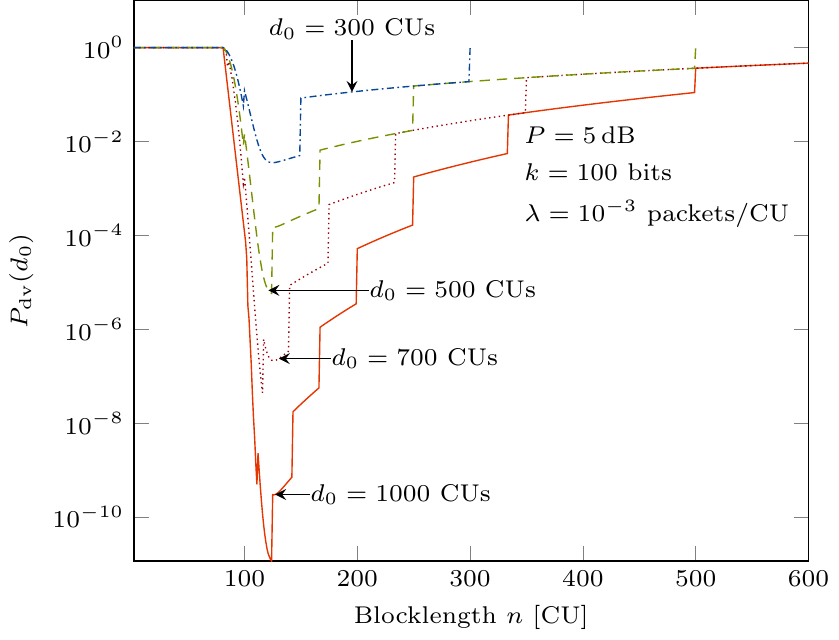}
\caption{Delay violation probability vs. blocklength.}\label{fig:delayviolation_vs_blocklength}
\end{figure}
\begin{figure}[t]
\centering
\includegraphics[width=.85\columnwidth]{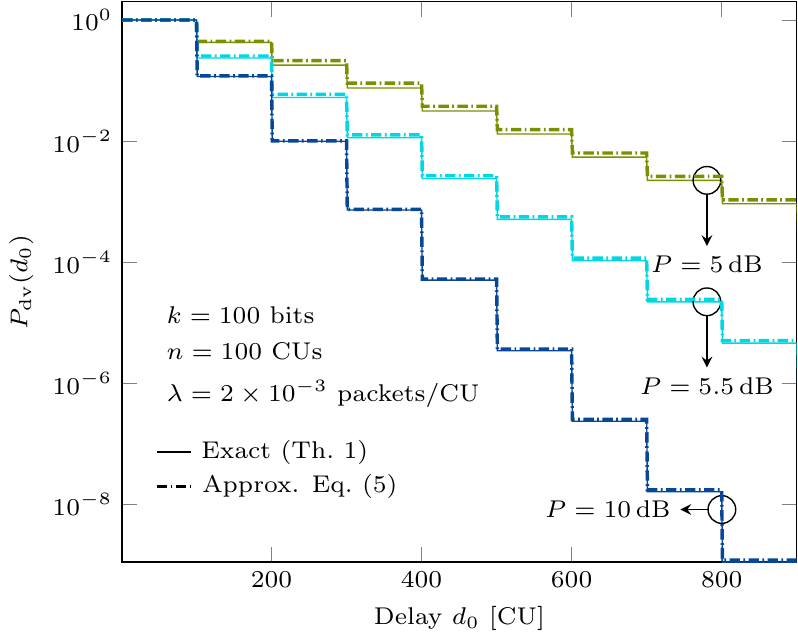}
\caption{Delay violation probability and its saddlepoint approximation.}\label{fig:delayccdf_compare_saddlepoint}
\end{figure}
\begin{figure}[t]
\centering
\includegraphics[width=.86\columnwidth]{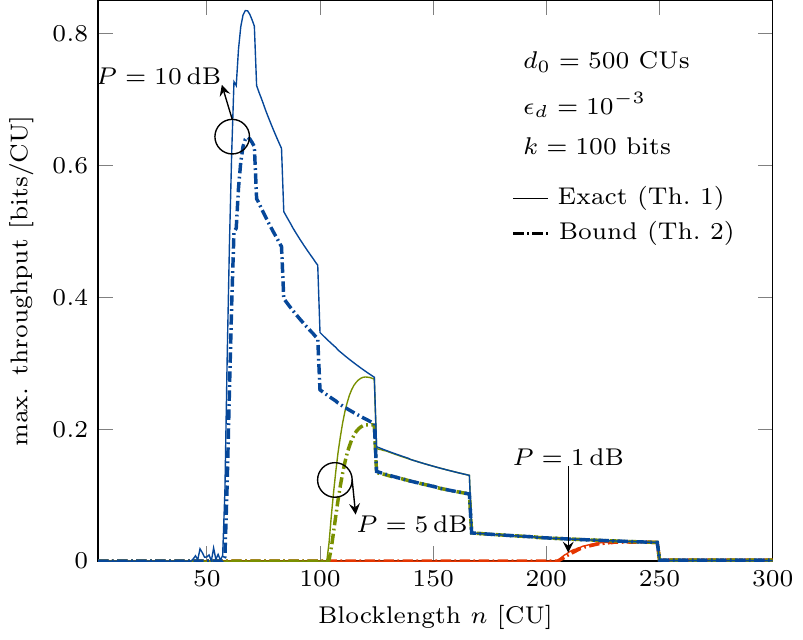}
\caption{Maximum throughput vs blocklength for three different SNR values.}\label{fig:maxarrivalrate_vs_blockLength_vary_snr_compare_netcalc}
\end{figure}
\begin{figure}[t]
\centering
\includegraphics[width=.85\columnwidth]{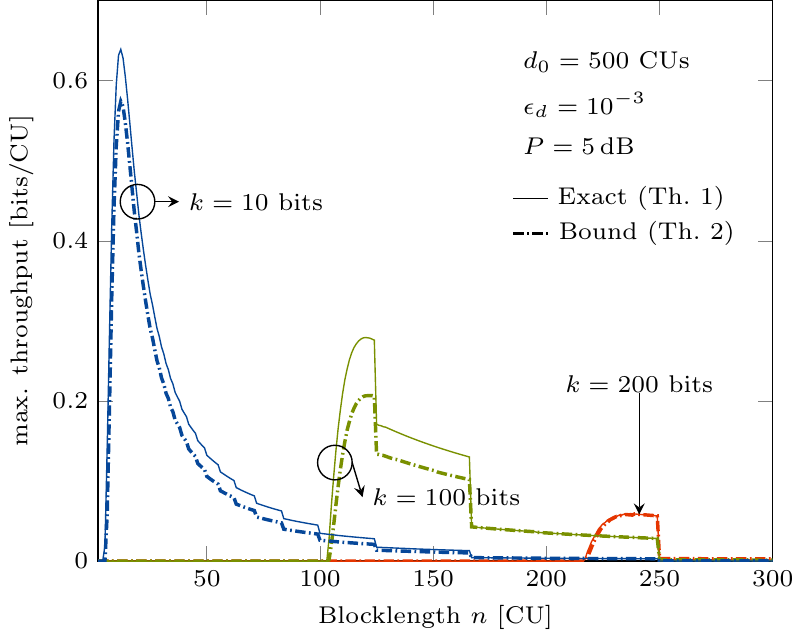}
\caption{Maximum throughput versus blocklength for three different  values of the information payload $\infobits.$}\label{fig:maxarrivalrate_vs_blockLength_vary_infobits_compare_netcalc}
\end{figure}
\section{Numerical Results}\label{sec:numerical_results}
Throughout this section, for a given size $\infobits$ (measured in bits) of the information packet  and for a given blocklength $\blocklength$, we use Shannon's achievability bound~\cite[Eq. (3)]{shannon1959probability}, which is the tightest achievability bound known for the AWGN channel,
to determine the packet error probability $\errorprob(\infobits,\blocklength)$.
For the parameter range considered in this section, using instead the easier-to-compute normal approximation~\cite[Eq.~(296)]{polyanskiy2010channel}  yields very similar results.

In Fig.~\ref{fig:delayviolation_vs_blocklength}, we illustrate the dependence of the delay violation probability on the blocklength $\blocklength$.
We choose $\snr=5\dB$, $\infobits=100$ information bits, and arrival rate $\averagearrivalrate=10^{-3}$ packets/CU. %
We observe there exists an optimum blocklength $\blocklength$, and, hence, an optimum code rate $\infobits/\blocklength$, that minimizes the delay violation probability for all  values of~$\delaythreshold$ considered in the figure.
In fact, on the one hand, when the blocklength is small, the packet error probability  is large and so is the number of retransmissions, yielding a large probability that the delay exceeds the threshold.
On the other hand, when the blocklength is large, the packet error probability is small, but even a small number of retransmissions is sufficient to generate a large delay.
Note that the jumps in the plot, which occur at submultiples of $\delaythreshold$, are caused by the change in the number of available retransmission rounds.

A comparison between the  delay violation probability~\eqref{expr:ccdf_steadystate_delay}, computed through a recursion based $z$-transform inversion, and the reduced-complexity saddlepoint approximation~\eqref{expr:ccdf_steadystate_delay_saddlepoint} is drawn in Fig.~\ref{fig:delayccdf_compare_saddlepoint}, where it is possible to appreciate that the saddlepoint approximation is extremely accurate.

Next, we study the maximum throughput, which we define as the product $k \averagearrivalrate^*$ between the number of information bits per packet $k$ and the maximum packet arrival rate $\averagearrivalrate^*$ achievable under a constraint on the delay violation probability.
In Figs.~\ref{fig:maxarrivalrate_vs_blockLength_vary_snr_compare_netcalc} and~\ref{fig:maxarrivalrate_vs_blockLength_vary_infobits_compare_netcalc}, the maximum throughput is plotted as a function of the blocklength $\blocklength$ for different values of $\snr$ and $\infobits$, respectively.
In both figures, we set $\delaythreshold=500$ CUs and a target delay violation probability $\delayviolationprob=10^{-3}$.
As a reference, we also plot throughput estimates obtained using the upper bound on the delay violation probability~\eqref{expr:ccdf_steadystate_delay_netcalc}, which relies on stochastic network calculus.
This bound is accurate only for  low SNR or large~$\infobits$.
Indeed, for the case $\snr=10\dB$ and $k=100$ bits depicted in Fig.~\ref{fig:maxarrivalrate_vs_blockLength_vary_snr_compare_netcalc}, the throughput estimate based on~\eqref{expr:ccdf_steadystate_delay_netcalc} is about $20\%$ off.
Although the bound~\eqref{expr:ccdf_steadystate_delay_netcalc} provides a loose throughput estimate, it predicts accurately the value of the throughput-maximizing blocklength.

To conclude, we elaborate on the difference between optimizing a system for a target average delay and optimizing it for a target delay violation probability.
To this end, we fix $\averagearrivalrate=10^{-3}$, ~$\snr=5\dB$, $\infobits=100$ bits, and  $\delaythreshold=500$ CUs.
For these parameters, the blocklength values of~$\blocklength=100$ CUs  and~$\blocklength=140$ CUs result in very similar average delays, namely about $154$ CUs and~$152$ CUs, respectively.
However, they yield significantly different delay violation probabilities, namely about $1.4\times 10^{-2}$ and~$2\times 10^{-4}$, respectively.
This highlights the importance of performing delay violation probability analyses in latency-critical wireless systems.

\begin{appendices}

  \section{Proof of Theorem~\ref{thm:delay_steadystate}}\label{proof:delay_steadystate}
  Let us denote by $\queuesizepktleave_m$ the number of bulks remaining in the buffer just after the $m$th bulk leaves the buffer, i.e.,
  \begin{IEEEeqnarray}{rCl}
  \queuesizepktleave_m &=& \queuesizeattime{\arrivaltimeofbulk{m}+\delayofpacket{m}}.\label{def:queuesize_packet_leave}
  \end{IEEEeqnarray}
  Since the number of bulks  arriving in the interval $(\arrivaltimeofbulk{m}+\delayofpacket{m},\arrivaltimeofbulk{m+1}+\delayofpacket{m+1})$ is independent of $\queuesizepktleave_m$, we conclude that $\curlybrac{\queuesizepktleave_m}_{m=1}^{\infty}$,  is a Markov chain governed by
  \begin{equation}
  \queuesizepktleave_{m+1} =
      \max\{\queuesizepktleave_m - 1,0\} + \sum_{t=1}^{\servicetimeofbulk{m+1}} \indicator{\bulkarrivalat{\arrivaltimeofbulk{m+1}+t}>0}.
  \end{equation}
  Note that the random variables $\curlybrac{\indicator{\bulkarrivalat{t}>0}}_{t=1}^{\infty}$ and $\curlybrac{\servicetimeofbulk{m}}_{m=1}^{\infty}$  are \iid, and independent of $\curlybrac{\queuesizepktleave_m}_{m=1}^{\infty}$.
  Hence,
  \begin{equation}
  \queuesizepktleave_{m+1} \sim
      \max\{\queuesizepktleave_m - 1,0\} + \arrivalsduringserving,
  \end{equation}
  where~$\arrivalsduringserving$ is the number of bulks of packets  arriving during the service time of a bulk, which is given by
  $  \arrivalsduringserving = \sum_{t=1}^{\servicetimeofbulk{1}} \indicator{\bulkarrivalat{t}>0}$.
The PGF of the steady-state buffer-size $\queuesizepktleave$ is~\cite[Eq. (11.3.11)]{grimmett2001probability}%
  \begin{IEEEeqnarray}{rCl}
  G_{\queuesizepktleave}(s) &=& \parantheses{1-\Bexpectation{\arrivalsduringserving}}\frac{(s-1)G_{\arrivalsduringserving}(s)}{s-G_{\arrivalsduringserving}(s)},\ \Bexpectation{\arrivalsduringserving}<1.\label{expr:pgf_steadystate_queuesize}
  \end{IEEEeqnarray}
Since $\queuesizepktleave \sim \sum_{t=1}^{\delay}\indicator{\bulkarrivalat{t}>0}$, then
$G_{\delay}(s) = G_{\queuesizepktleave}({G_{\indicator{\bulkarrivalat{1}>0}}^{-1}\parantheses{s}})$, where
  \begin{IEEEeqnarray}{rCl}
  G_{\indicator{\bulkarrivalat{1}>0}}\parantheses{s} &=& (1-\packetarrivalprob)^\blocklength + s\parantheses{1-(1-\packetarrivalprob)^\blocklength}. \label{expr:pgf_bulkarrival_greatherthanzero}
  \end{IEEEeqnarray}
Furthermore, from the definition of $\arrivalsduringserving$ and from~\eqref{def:servicetime_bulk}, we obtain the equality
  \begin{IEEEeqnarray}{rCcCcl}
  \Bexpectation{\arrivalsduringserving} &=& \Bexpectation{\indicator{\bulkarrivalat{1}>0}}\Bexpectation{\servicetimeofpacket{1}}\Bexpectation{\bulkarrivalcountat{1}}
&=&\packetarrivalprob\blocklength/(1-\errorprob).
  \end{IEEEeqnarray}
  Next, we observe that
$G_{\arrivalsduringserving}(s) =
    G_{\servicetimeofbulk{1}}\parantheses{G_{\indicator{\bulkarrivalat{1}>0}}\parantheses{s}}$,
    $G_{\servicetimeofpacket{1}}\parantheses{s} = \parantheses{1-\errorprob} s/(1-\errorprob s)$, %
and $G_{\servicetimeofbulk{1}}\parantheses{s}
  = G_{\bulkarrivalcountat{1}}\parantheses{G_{\servicetimeofpacket{1}}\parantheses{s}}$ %
  where
  \begin{IEEEeqnarray}{rCl}
  G_{\bulkarrivalcountat{1}}\parantheses{s} &=& \frac{\parantheses{1-\packetarrivalprob+\packetarrivalprob s}^\blocklength-\parantheses{1-\packetarrivalprob}^\blocklength}{1-\parantheses{1-\packetarrivalprob}^\blocklength}. \label{expr:pgf_bulkarrivalcount}
  \end{IEEEeqnarray}
  Algebraic manipulations yield~\eqref{expr:pgf_steadystate_delay}.%
%
\end{appendices}

\linespread{0.90}
\bibliographystyle{IEEEtran}
\bibliography{IEEEabrv,publishers,confs-jrnls,refs}

\end{document}